Chapitre 6

# Modèle FBS-PPR : des objets d'entreprise à la gestion dynamique des connaissances industrielles

## 6.1. Introduction

Les étapes du cycle de vie d'un produit industriel peuvent être décrites sous la forme d'un réseau de processus métier. Les produits matériels et informationnels sont à la fois matières premières et résultats de ces processus. Une valorisation aussi complète que possible des connaissances passe par des efforts de modélisation en mettant l'accent sur la complétude et la généricité. Seule une démarche de standardisation impliquant plusieurs domaines, tels que la modélisation des produits, la modélisation des processus, la modélisation des ressources et l'ingénierie des connaissances, peut permettre de construire un système documentaire plus rationnel et plus rentable.

---

Chapitre rédigé par Michel LABROUSSE, Nicolas PERRY et Alain BERNARD.

## 6.2. Contexte industriel / Besoin de standardisation

### 6.2.1. Gérer l'information, un enjeu majeur pour les entreprises

Les entreprises ont longtemps porté leurs efforts d'optimisation sur les phases de production et de gestion des stocks : réduction du nombre de composants gérés par l'entreprise, réduction des stocks et en-cours, rationalisation des processus, mécanisation des phases de fabrication et d'assemblage, etc.

Néanmoins, il est vite apparu que les phases d'ingénierie impactaient considérablement la productivité et la compétitivité des entreprises. Les coûts ultérieurs du cycle de vie du produit découlent directement des choix effectués lors de la conception : procédés et technologies de fabrication, aptitude à un assemblage rapide et économique, coûts du service après vente, etc. De certains choix effectués en conception dépendront aussi les possibilités d'apporter à un produit des évolutions ultérieures - qu'elles soient techniques ou cosmétiques - pour le remettre au goût du jour.

Dans un souci d'amélioration de la phase de conception, les différents services de l'entreprise ont alors commencé à se décloisonner : des équipes projets ont été créées et les sous-traitants ont de plus en plus été impliqués dès les premières phases de conception. Des retours d'expérience et des savoir-faire « métiers » ont ainsi pu être mieux pris en compte. Cela a aussi permis de réduire les coûts et les temps de développement des nouveaux produits, vecteurs importants du maintien d'une position concurrentielle pour les entreprises.

Cette mise en commun d'informations hétéroclites - bien que très bénéfique - a vite révélé une gestion des connaissances assez peu satisfaisante : les systèmes conçus sont parfois d'une extrême complexité et les réunions projets se révèlent extrêmement coûteuses (plusieurs personnes souvent très qualifiées sont mobilisées) et ne conduisent que trop rarement à de réelles avancées. Faute de référentiels, de règles et d'indicateurs de performance fiables, les interlocuteurs en présence campent sur leur position, ne parviennent pas toujours à un réel consensus, sans compter que les débats s'enveniment parfois.

Une gestion plus pertinente de ces informations, contribuant plus largement à celle des connaissances et des compétences, apparaît donc comme un enjeu majeur pour les entreprises.

Une contribution significative allant dans ce sens est relative à un référentiel numérique, commun à tous les intervenants et présenté en figure 6.1.

Facilitant l'ingénierie concourante, il permet donc à chaque service d'intervenir sur la conception avant que les choix effectués ne deviennent coûteux ou irréversibles. Le référentiel numérique permet également une réutilisation de l'existant, c'est à dire des connaissances précédemment capitalisées. Cela suppose une structuration cohérente des données liées au produit (modèle structurel, modèle fonctionnel, modèle physique), ainsi qu'aux processus de son cycle de vie.

Les enjeux du référentiel numérique (cf. Fig. 6.1.) sont donc particulièrement attractifs puisqu'ils permettent la réduction des temps de mise sur le marché des produits ainsi que la baisse des coûts de développement et de fabrication, tout en assurant la maîtrise de la qualité.

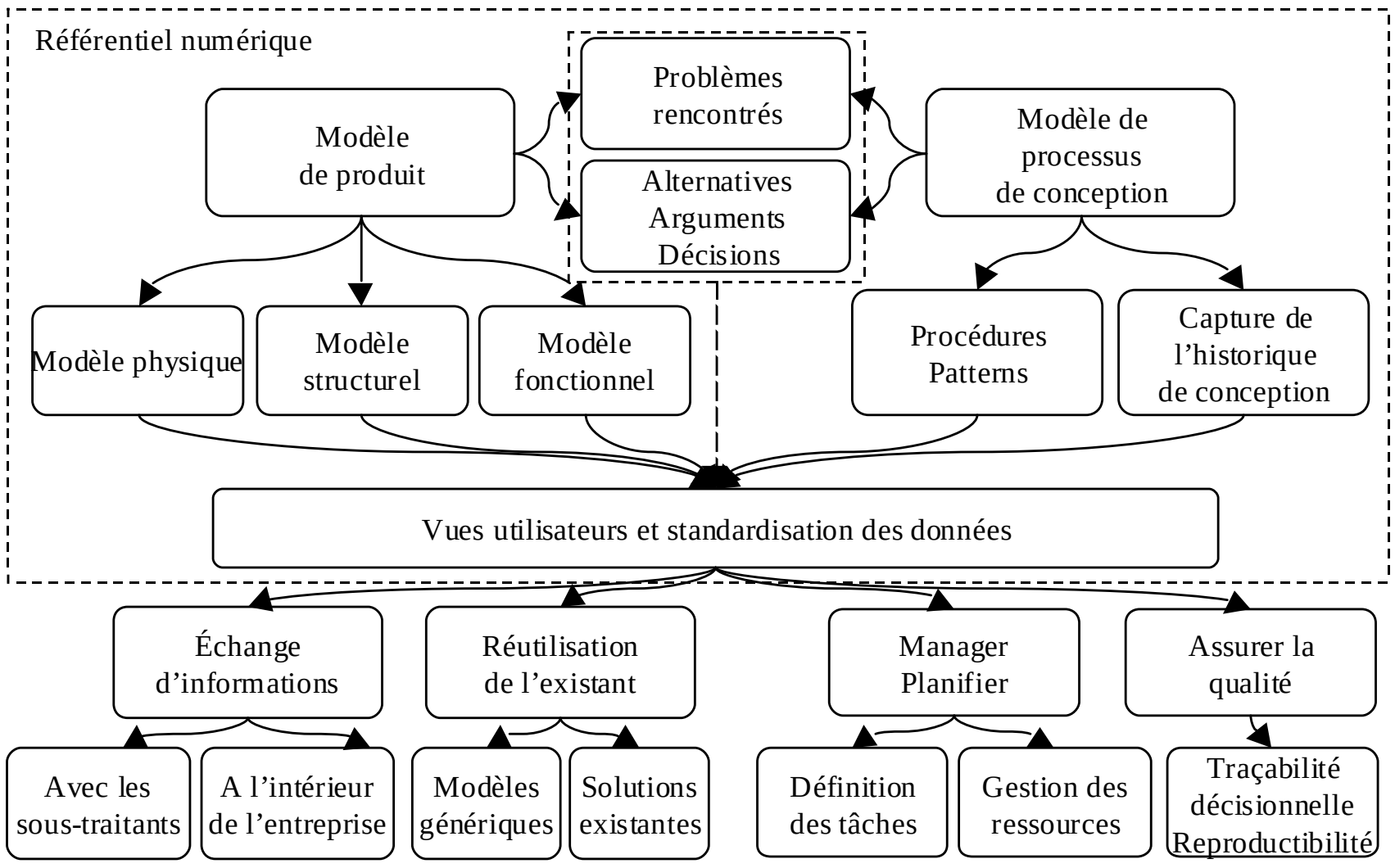

**Figure 6.1.** *Contenu et enjeux du référentiel numérique*

Le référentiel numérique contient des informations structurées ainsi que des informations enfouies dans des documents semi ou non structurés supports de connaissance. La constitution de ce référentiel ne se limite pas à un simple stockage de contenus, mais doit intégrer des problématiques de capitalisation, d'indexation et de réutilisation pertinente de ces connaissances. En particulier, des difficultés de représentation inhérentes aux concepts de connaissance et de compétence sont à analyser.

### 6.2.2. Difficultés liées aux concepts de connaissance et de compétence

Pour Michel Grundstein [GRU 00], la connaissance peut être caractérisée par trois postulats :

– postulat 1 : la connaissance n'est pas un objet, elle résulte de la rencontre d'une donnée avec un sujet, et s'inscrit au travers du système d'interprétation de l'individu dans sa mémoire.

– postulat 2 : la connaissance est reliée à l'action. Du point de vue de l'entreprise, la connaissance est crée par l'action et est essentielle à son déroulement. Elle est finalisée par l'action.

– postulat 3 : il existe deux grandes catégories de connaissances de l'entreprise : les éléments tangibles (connaissances formalisées) et les éléments intangibles (connaissances incarnées par des personnes).

Ces postulats soulèvent les discussions suivantes :

– au sens strict, seule l'information peut être formalisée et donc capitalisée : rien ne permet d'affirmer a priori que différents utilisateurs interpréteront des informations de la même manière. Néanmoins, si les informations sont destinées à une catégorie précise d'utilisateurs, c'est à dire ayant des formations, des expériences et donc des manières semblables de penser, on peut supposer que les informations seront interprétées de manière similaire et donc les assimiler à des connaissances.

– la connaissance est indissociable de l'action. La connaissance est de l'information contextualisée : elle suppose la définition de l'étendue de validité de l'information, de critères d'applicabilité, etc. En resituant l'information dans son contexte (utilisateur / domaine), les variations d'interprétation peuvent être considérablement réduites. Ainsi, par la définition précise du triplet « information/utilisateur/domaine », l'information peut s'apparenter à de la connaissance.

– en fait, on peut considérer qu'il existe trois catégories de connaissances : les savoirs formalisés (ou explicites), les savoirs formalisables et les savoirs tacites (ou intangibles). Alsène [ALS 03] définit ainsi ces trois catégories :

Les **savoirs formalisés** (ou explicites) sont « les savoirs qui ont déjà été explicités par l'intermédiaire d'un langage rationnel, que l'on peut retrouver dans des livres, des plans, des manuels de procédures, des systèmes d'information, des bases de connaissances, etc. » ;

Les **savoirs formalisables** sont « les savoirs qui n'ont pas encore été formalisés, explicités, officialisés au moyen d'un langage rationnel, mais qui pourraient l'être (par exemple, certains savoir-faire développés par des individus ou des communautés de pratique, ou encore, certains savoirs informels, implicites, concernant le contexte social du travail) » ;

Les **savoirs tacites** (ou intangibles) sont « les savoirs qui ne peuvent pas être explicités par l'intermédiaire d'un langage rationnel, et qui par conséquent sont très difficilement transmissibles d'un individu à l'autre, sinon par l'observation, l'imitation, la socialisation et le recours à la métaphore (par exemple, certains savoir-faire, mais aussi certains schémas mentaux) ».

Finalement, la **connaissance** peut être définie comme le résultat d'une interaction entre des informations et un système d'interprétation dans un domaine d'application donné. Elle peut être caractérisée par le triplet « information/utilisateur/domaine ».

La **compétence** nécessite la mise en œuvre de connaissances. Il s'agit d'une capacité et d'une habileté à exploiter des connaissances et des ressources pour exercer une activité dans un contexte contraint donné et atteindre un objectif [LAB 03].

Un schéma de synthèse est proposé par Gardoni [GAR 99] (cf. Fig. 6.2.) :

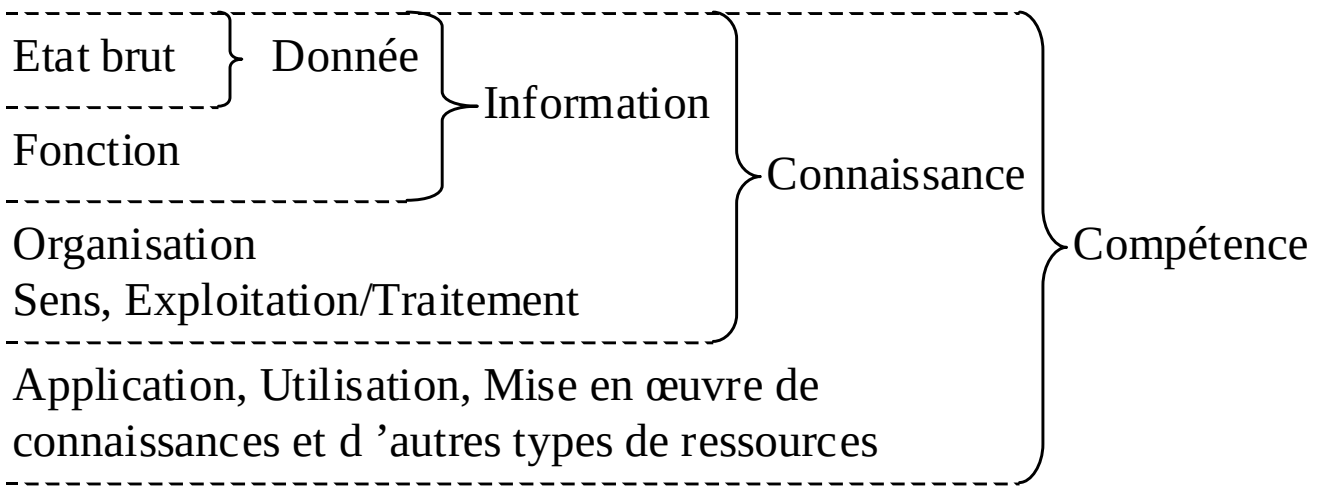

**Figure 6.2.** *Lien entre connaissance et compétence*

Cette structuration et cette caractérisation de la connaissance constituent une simplification de la réalité qui comporte les inconvénients suivants :

– même en spécifiant une catégorie d'utilisateur donnée, les processus d'interprétation ne pourront jamais être rigoureusement les mêmes. Ces processus cognitifs sont en effet sensibles à beaucoup trop de paramètres (l'ensemble du vécu de l'utilisateur, ses caractéristiques neurologiques, ses pensées et préoccupations actuelles, son état de fatigue, etc.) ne sont pas toujours déterministes et ne peuvent pas être modélisés : le système d'interprétation d'un individu n'est pas totalement explicitable.

– la transmission de connaissances est un processus complexe. Pour être efficace, l'apprentissage suppose une très grande cohérence dans le choix et l'ordre de présentation de l'information (la difficulté de définition des programmes de l'Education Nationale et leurs fréquentes évolutions en sont un exemple) : l'acquisition d'un nouveau savoir suppose des pré-requis qu'il est important de définir et de transmettre préalablement. Le sujet ne peut acquérir des connaissances que progressivement ; les processus d'interprétation se construisent au fur et à mesure (cela est particulièrement vrai pour certaines sciences comme les mathématiques, pour lesquelles les axiomes s'apparentent vite à des évidences… pourtant bien arbitraires). Des exemples applicatifs et des exercices à la présentation « formatée » suivant des conventions partagées participent grandement à ce processus d'apprentissage.

Ainsi, modéliser des connaissances est d'une extrême complexité. En plus de l'information véhiculée, il faut amener le sujet à penser, à interpréter les informations de manière similaire. Il est donc très important de définir le contexte et d'expliciter les connaissances requises pour l'interprétation de ces nouvelles connaissances. Ces efforts passent par une standardisation à plusieurs niveaux, de la représentation des informations, de leur diffusion et de leur présentation à l'utilisateur ainsi que de la normalisation nécessaire à une interprétation homogène des connaissances.

### 6.2.3. Besoin de standardisation

De nos audits réalisés en milieu industriel ressort une nécessité de standardisation :

– les besoins d'échange entre partenaires et entre services sont nombreux. L'interdépendance de différents métiers de l'entreprise ou en sous-traitance rend nécessaire l'échange de fichiers et de documents. Des formats souvent propriétaires (CAO, FAO, etc.) apparaissent donc comme un obstacle à la communication : les logiciels ne peuvent être installés sur les postes de tous les intervenants, et même si des visualisateurs (« viewers ») existent parfois, ils nécessitent néanmoins d'être préalablement installés et requièrent parfois des formations spécifiques pour être utilisés à bon escient. De ce fait, le format papier demeure assez fortement employé, avec des risques d'obsolescence non négligeables.

– ces échanges seraient grandement simplifiés par une plus grande interopérabilité entre logiciels. En effet, les sous-traitants et donneurs d'ordres ne sont pas toujours équipés des mêmes logiciels. Les échanges sont alors basés sur des formats neutres impliquant parfois des conversions approximatives ou avec pertes. Ainsi des surfaces échangées au format IGES seront-elles qualifiées de « mortes » et les possibilités de modification seront alors très réduites. La définition de formats d'échange communs et leur intégration dans les logiciels apparaissent donc comme un enjeu majeur pour les entreprises. Mais les leaders du marché du logiciel ont souvent des logiques anticoncurrentielles, et préfèrent donc rendre captive leur clientèle.

– au-delà de l’échange, il existe des difficultés potentielles de compréhension, de traitement de l’information. En effet, pour que l’information puisse générer des connaissances, un traitement de l’information est nécessaire. Pour rendre les informations interprétables (par des hommes ou des logiciels) une standardisation des informations et des processus associés est généralement requise. La dualité information / connaissance explique aussi de nombreuses initiatives pour modéliser la connaissance : modèles multi-utilisateurs [BLA 98], modèles multi-représentations [TIC 95], vues métiers [HAR 97], etc. Pour que l’information génère des connaissances, elle sera traitée, filtrée et présentée différemment selon le type d’utilisateur auquel elle s’adresse.

Afin d’améliorer la formalisation, la gestion et la réutilisation des connaissances, une modélisation plus générique ainsi que son déploiement sur les trois vues processus, produit et ressource va être proposée après une synthèse sur les modélisations de type FBS (Function, Behavior, Structure),.

## 6.3. Modélisation FBS (Fonction/Comportement/Structure)

### 6.3.1. Fonction, comportement et structure

Le modèle Function, Behavior, Structure (FBS) (cf. figure 6.3.) est une approche pour concevoir des produits, qui permet la représentation de manière explicite des fonctions du produit (le problème), de la structure du produit (la solution) et des comportements internes du produit [HU 00].

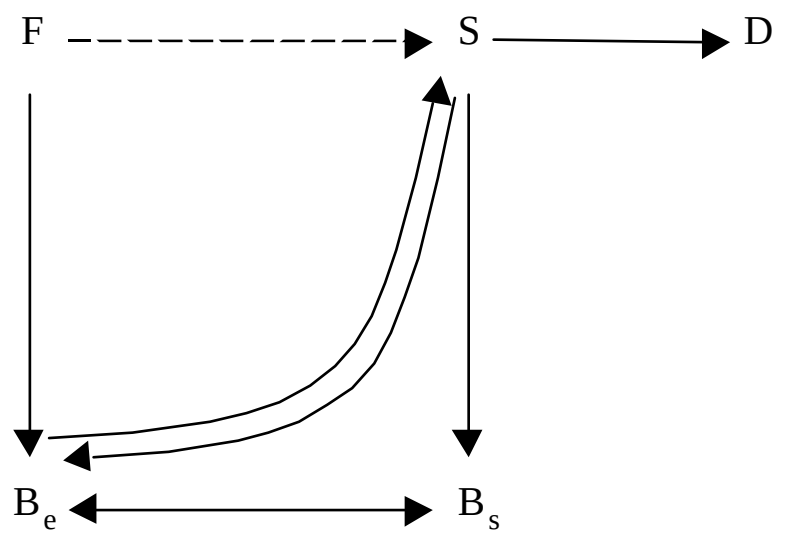

**Légende :**

| | |
|---|---|
| F | Function |
| Be | Expected Behavior |
| S | Structure |
| Bs | Structure Behavior |
| D | Design Description |

**Figure 6.3.** *Modèle FBS proposé par Gero [GER 90]*

### 6.3.2. Notions complémentaires de FBS

L'étude du milieu industriel [BER 02] et l'analyse bibliographique effectuées amènent au fait que le modèle FBS constitue une première base utilisable pour notre approche. Néanmoins il doit être enrichi pour pouvoir modéliser convenablement les processus d'entreprise et leurs composants. Les aspects complémentaires suivants doivent être pris en compte :

– la notion d'indicateurs de performance (PI) : la définition d'indicateurs de performance est rendue nécessaire pour évaluer les solutions et effectuer des choix. Ces indicateurs peuvent être de plusieurs natures : économiques (coût d'acquisition, à l'usage, etc.), mécaniques (rigidité, état de surface, etc.), esthétiques (surfaces de classe "A", finitions, etc.), temporels (durée d'allumage, temps de cycle, durée de vie, etc.), etc. Dans le modèle proposé ci-après, il est intéressant de noter que ces indicateurs sont adimensionnels (ratio réalisé/attendu).

– la notion d'état : la notion d'état est présente implicitement dès lors que l'on fait intervenir des processus. Les changements d'états peuvent être considérés comme la représentation d'un comportement.

– la notion de processus et de contexte : dans [MAG 98] est proposé le modèle FBPS (Function/Behavior/Process/Structure). La notion de comportement émerge de l'application d'un processus à une structure. Par exemple lors de la fabrication d'un produit, le comportement pourra être défini comme le résultat économique, temporel ou de qualité d'un processus. Un aspect intéressant est que le comportement est ici contextualisé, puisqu'il est le résultat d'un processus donné.

– la notion d'effets externes, d'environnement et de contraintes [WAN 02], [DEN 00] : l'aspect contextuel peut être amélioré par la prise en compte de l'environnement externe et de ses interactions avec le système.

– la notion de ressource : il est nécessaire de distinguer la structure résultante (le produit) et la structure permettant d'atteindre cet objectif c'est à dire les ressources mises en œuvre. Or le modèle FBS "classique" ne prend pas en compte les ressources (matérielles, logicielles ou humaines). Celles-ci jouent pourtant un rôle prépondérant durant les processus de conception, fabrication, etc.

En prenant en compte les remarques précédentes, la modélisation FBS-PPR a pu être définie.

## 6.4. Proposition pour une modélisation unifiée : le modèle FBS-PPR

Le modèle FBS-PPR (Function / Behavior / Structure - Process / Product / Resource) consiste en le déploiement du modèle FBS suivant trois vues : la vue processus, la vue produit et la vue ressource.

Dans la modélisation FBS-PPR, le produit est défini comme le résultat, l’objet d’un processus. Les ressources (matérielles, logicielles, énergétiques ou humaines) sont des moyens qui participent au processus mais qui, contrairement au produit, n’en sont pas la finalité. Les processus (de conception, de fabrication, de contrôle, etc.) sont une organisation temporelle, spatiale et hiérarchique d’activités qui font appel à des ressources (ou moyens) et qui conduisent à des produits (ou sorties).

– objet d’entreprise

Dans le but d’avoir un élément générique englobant les notions de processus, de produit et de ressource, est définie la notion d’*objet d’entreprise*. Un objet d’entreprise est une entité constitutive de l’entreprise et/ou manipulée par elle et qui joue un rôle dans son fonctionnement. Cet élément générique est destiné à permettre une gestion plus homogène et donc plus efficiente des notions de processus, de produit et de ressource.

– rôles de processus, de produit et de ressource

Une hypothèse forte du modèle FBS-PPR est que les notions de processus, de produit et de ressource sont des notions abstraites et circonstancielles, ce qui implique qu’un objet ne peut pas être considéré comme étant durant tout son cycle de vie soit un processus, soit un produit, soit une ressource.

En effet, supposons qu’un concepteur dessine une pièce en CAO. Ce plan CAO sera le produit de son étude. Mais ce produit servira de ressource à un programmeur pour la création de la gamme d’usinage. Ce processus servira alors de ressource à l’employé qui fabriquera la pièce sur la machine-outil.

Ainsi on s’aperçoit qu’un objet d’entreprise (pouvant être matériel ou non) peut être tour à tour processus, produit ou ressource. On ne peut donc pas dire d’un objet d’entreprise qu’il est, mais plutôt qu’il joue un rôle circonstanciel de processus, de produit ou de ressource. Cette distinction dans la modélisation va, du moins partiellement, à contresens de l’usage, pour lequel la notion de produit renvoie le plus souvent à un objet matériel destiné à la vente.

### 6.5. Dynamique et comportements dans le modèle FBS-PPR

Si les notions de fonction et de structure ne posent pas de problème majeur - elles sont des parties constitutives de l’objet, il n’en va pas de même pour le comportement qui est contextuel et ne peut pas être défini indépendamment d’un processus (celui-ci permet en particulier de délimiter le domaine d’application d’un comportement).

Le comportement n’est pas intrinsèque à l’objet, il résulte d’une interaction entre celui-ci et son environnement via des stimuli. En fait, il ne faudrait donc pas parler du comportement, mais des comportements de l’objet. Les processus apportent en effet l’aspect temporel (donc la dynamique) et l’aspect contextuel aux comportements des objets.

La gestion de la notion de comportement est une originalité et un apport du modèle FBS-PPR. Pour la rendre plus explicite, une représentation graphique est proposée en Figure 6.4.

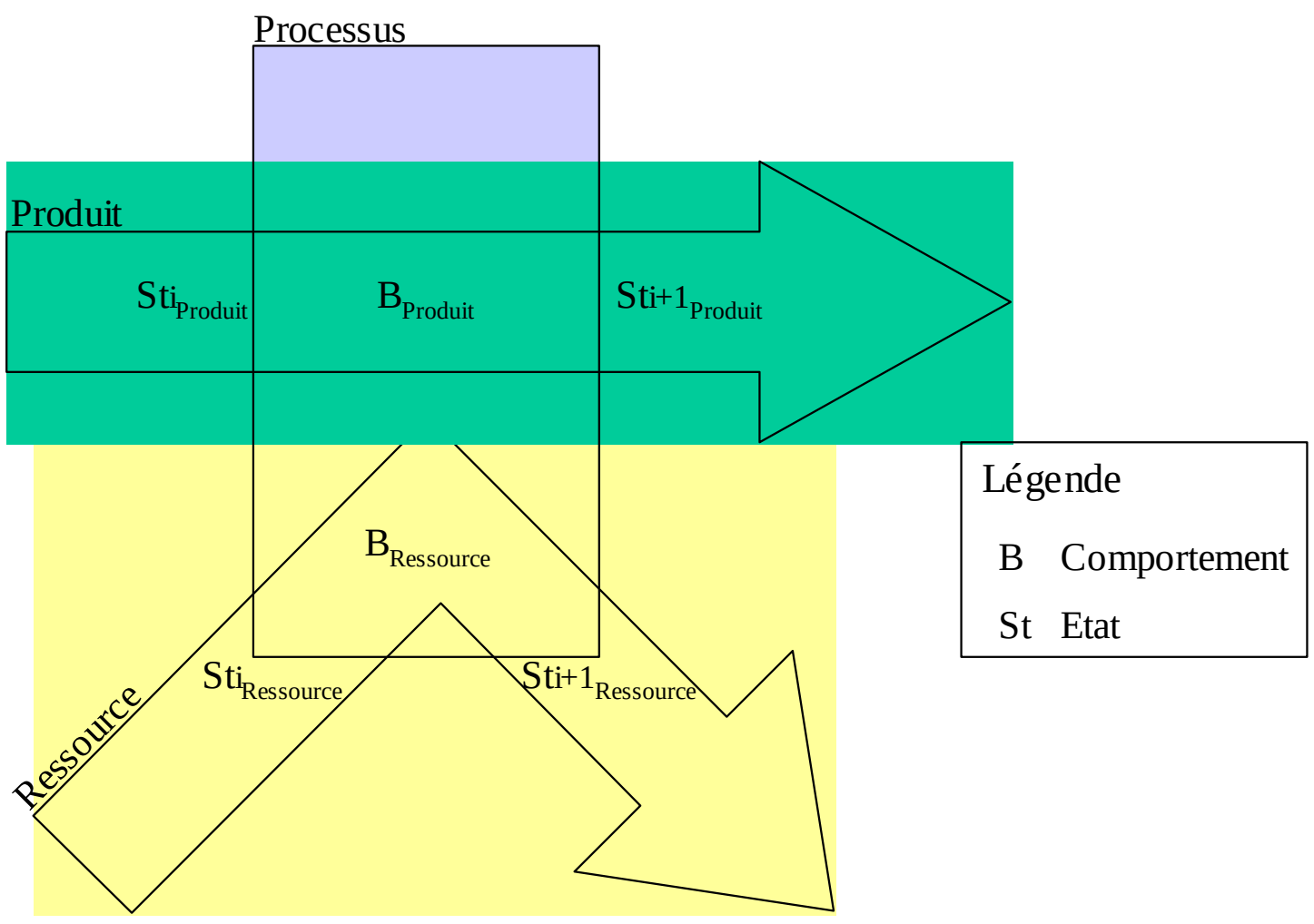

**Figure 6.4.** *Gestion de la notion de comportement dans le modèle FBS-PPR*

Dans un souci de simplicité d’écriture, un « objet jouant le rôle de processus » sera qualifié dans la suite d’« élément processus » et respectivement pour les objets jouant un rôle de produit ou de ressource d’éléments produit ou ressource.

L’élément processus (au centre de la représentation) joue un rôle prépondérant dans le modèle FBS-PPR :

– il permet d’une part de relier l’ensemble des éléments intervenant au cours d’une activité donnée et donc de définir une partie du contexte ;

– il permet d’autre part de gérer l’évolution des éléments produits et ressources au travers de leurs changements d’états.

Les états traduisent des modifications de la structure et représentent les entrées sorties de l'élément processus. Ils permettent de caractériser le comportement des objets, l'évolution de leurs structures.

Ce comportement peut être considéré comme une suite d'états discrets [UME 90] :

B = (St1, St2, …, Sti, …, Stn)

Bien entendu, cette suite d'états n'est pas entièrement représentative du comportement, puisqu'elle se limite à une vue discrète des phénomènes. Des lois physiques (modèles continus) peuvent permettre d'expliquer ces changements d'états successifs. Les relations entre comportement et états sont alors nommées relations B-S [UME 90].

Sur ces bases, dans le modèle FBS-PPR, le comportement comprend des états, des déclencheurs et des lois de comportement. La notion d'état renvoie à des « instantanés » de la structure : la notion d'état est donc une passerelle entre comportement et structure.

Supprimé :

Remarque : il n'a sans doute pas échappé au lecteur que la Figure 6.4. n'explicite pas la gestion du comportement de l'élément processus. En fait, le comportement d'un élément processus, donc de ses états et statuts (en cours, en attente, terminé, etc.), ne peut être modélisé que vis à vis d'un autre élément processus pour lequel il sera vu comme un élément produit. Il a donc été nécessaire de créer un élément processus abstrait nommé « déroulement du temps » n'ayant pas de comportement (sinon le problème se poserait à l'infini !) et permettant de caractériser le déroulement d'un élément processus donné au cours du temps.

### 6.6. Représentation UML simplifiée du modèle FBS-PPR

Bien que pouvant jouer des rôles différents (processus, produit ou ressource), les objets d'entreprise sont modélisés suivant le même méta-modèle formalisé en UML (cf. Fig. 6.5.). La classe centrale du modèle FBS-PPR est la classe *Objet*.

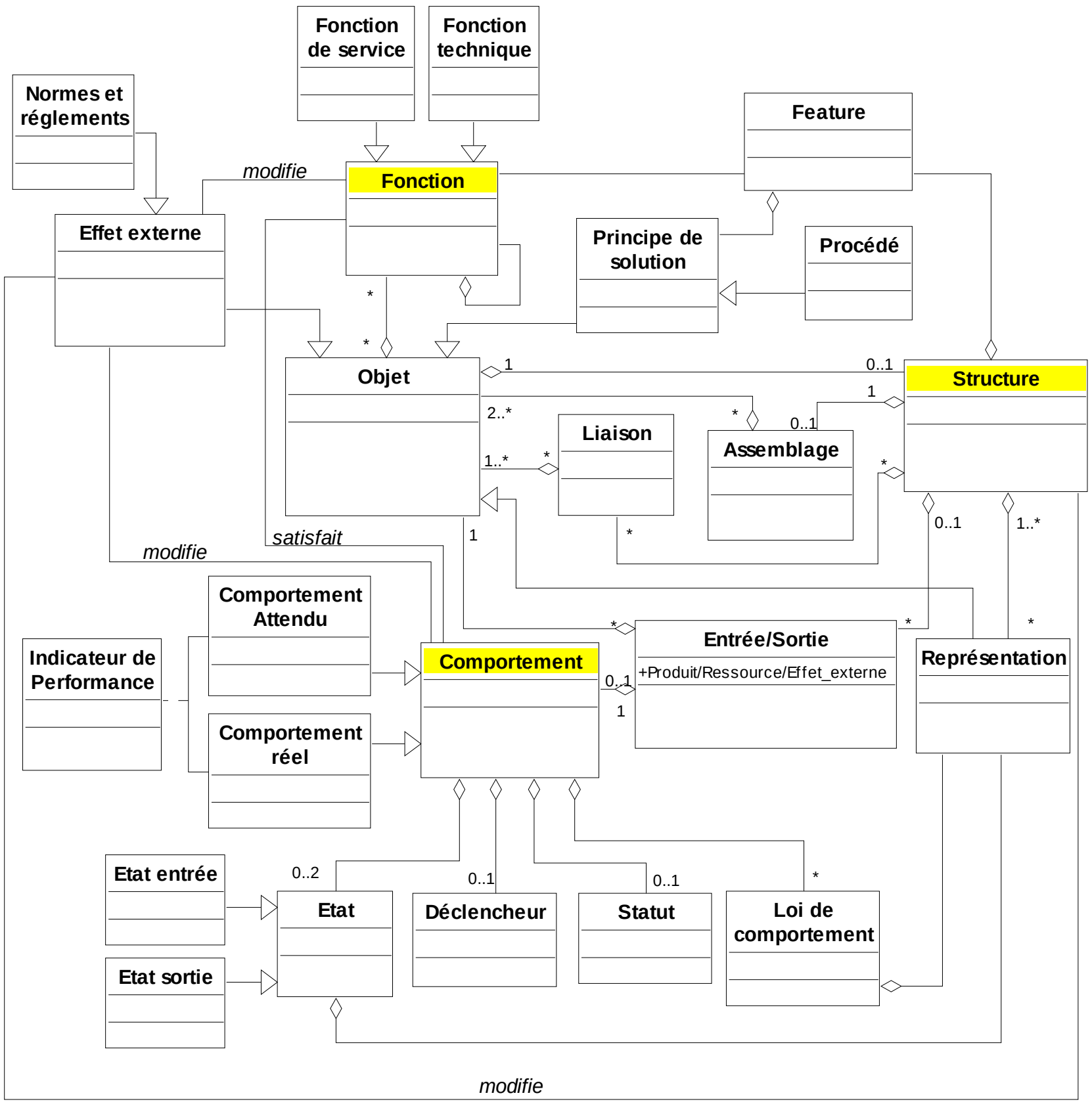

**Figure 6.5.** *Représentation UML simplifiée de la notion d'objet d'entreprise*

Les objets disposent tous d'un modèle structurel : un objet peut se décomposer en sous-objets organisés hiérarchiquement et/ou séquentiellement (utilisation des classes *Liaison* et *Assemblage*). Par exemple, une voiture peut être décomposée en sous ensembles (moteur, châssis, etc.) qui se décomposeront à leur tour jusqu'aux pièces et composants (vis, écrous, pneus, etc.)

La classe *Représentation* permet des descriptions statiques de l'objet : par exemple des fichiers CAO, de calcul, de coûts, etc.

La classe *Entrée/Sortie* permet de gérer les entrées et sorties de l'objet, et sera donc particulièrement utile pour expliciter les produits et ressources d'un élément processus (les entrées et sorties sont des élément structurels d'un objet temporel).

Comme cela a été détaillé (cf. paragraphe 6.5.), le comportement n'est pas directement reliée à l'objet (produit ou ressource) puisqu'il résulte de l'interaction entre cet objet et un élément processus : la classe *Entrée/Sortie* permet donc de relier l'objet à un objet jouant le rôle de processus tout en y associant les *Comportements.*

Les *Comportements* - attendus ou réalisés - font appel à des états (par exemple une pièce dans son état brut et dans sont état usiné) qui représentent la part discrète des phénomènes, à des déclencheurs, et à des équations de comportement représentatives de la partie continue des phénomènes (par exemple des équations différentielles).

La classe *Indicateur de performance* permet de comparer un comportement attendu et un comportement réel et donc d'évaluer le taux de satisfaction (vecteur d'écart) des objectifs.

La classe *Fonction* permet de modéliser les fonctions de l'objet, qui peuvent être décomposées en sous-ensembles : par exemple les fonctions de service et les fonctions techniques.

Les classes *Feature* et *Principe de solution* permettent de conserver des éléments de solution réutilisables (souvent appelés « pattern » en anglais) [GZA 00] : par exemple des formes usuelles (clavette, support de fixation normalisé, etc.) accompagnées de processus optimisés pour leur réalisation (usinage, fonderie, etc.).

La classe *Effet externe* permet de modéliser des contraintes ayant une influence sur l'objet considéré. Par exemple des influences géopolitiques (variations brutales du cours de matières premières, ouverture ou fermeture de marchés, évolution de taxes douanières, dévaluation de monnaies, etc.), des évolutions dans la stratégie de l'entreprise suite à un changement de direction ou encore les lois, règlements et normes en vigueur.

## 6.7. Principaux apports du modèle FBS-PPR

### 6.7.1. Une meilleure homogénéité et complétude du modèle

En FBS-PPR, les informations concernant un objet (quel qu'il soit) sont classées suivant trois classes de base : fonction, comportement et structure.

Le modèle est donc beaucoup plus facile à appréhender qu'un modèle ou les noms de classes diffèrent selon le type d'objet considéré (par exemple certaines modélisations utilisent la terminologie fonction lorsque l'on considère un produit et celle d'objectif lorsque l'on considère un processus).

L'approche FBS-PPR offre aussi une bien meilleure complétude que les approches FBS usuelles qui ne permettent en général que la modélisation FBS du produit (mais pas de manière conjointe des ressources et des processus). Un exemple d'instanciation de FBS-PPR est proposé en Figure 6.6. : il illustre bien le fait que les éléments produits et ressources sont définis relativement à chaque élément processus et non relativement au processus global.

| FBS-PPR | | Processus | Ressource | Produit |
|---|---|---|---|---|
| Concevoir liaison bâti/sol | Déf. alternatives | Définir des alternatives pour la liaison du bâti/sol. | Concepteur BE<br>Logiciel de CAO<br>Documentation fournisseurs | S1: 4*(vis+rondelle) & bâti avec 4 trous.<br>S2: 4*(goujon,écrou+rondelle) & bâti avec 4 trous. |
| | Evaluation alternatives | Evaluer les alternatives de solution pour la liaison bâti/sol. | Concepteur BE<br>Logiciel CAO et calcul<br>Documentation fournisseurs | S1: Fonction OK, coût faible, démontages fréquents : non.<br>S2: Fonction OK, côut moyen, démontage et remontages OK. |
| | Choix de solution | Choisir la solution la plus économique répondant au besoin | Concepteur BE<br>Logiciel de CAO<br>Documentation fournisseurs | S1: retenue car répondant au besoin et faible coût.<br>S2: non retenue. Fonctions supplémentaires inutiles. |

| FBS-PPR | | Processus | Ressource | Produit |
|---|---|---|---|---|
| Perçage 4 trous ∅10 sur le bâti | Montage outil (foret ∅10) | Equiper la perceuse d 'un mandrin avec foret ∅10. | OS<br>Mandrin et foret ∅10<br>Outils d 'atelier<br>Plan de fabrication | Perceuse |
| | Bridage du bâti sur la perceuse | Brider le bâti sur la perceuse. | OS<br>Perceuse<br>Brides<br>Outils d'atelier | Bâti non percé |
| | Réglage de la perceuse | Régler la perceuse. | OS<br>Plan de fabrication | Perceuse avec bâti bridé |
| | Perçage d'un trou | Percer un trou ∅10. | OS<br>Perceuse<br>Plan de fabrication | Bâti |
| | Débridage de la pièce | Débrider la pièce. | OS<br>Outils d'atelier | Perceuse avec bâti bridé |

**Figure 6.6.** *Exemple d'instanciation de la brique FBS-PPR [LAB 03]*

### 6.7.2. Gestion PPR innovante via la notion d'objet d'entreprise et de rôle

La gestion disjointe des espaces produit, processus et ressource que l'on retrouve dans la plupart des modélisations trouve sans doute son origine dans des héritages historiques et le poids des logiciels établis : gestion des produits (CAO) d'une part, des ressources d'autre part (planification de ressources, implantation de machines, gestion de production) et enfin des processus (management de projet, processus de fabrication).

En FBS-PPR, la gestion des objets d’entreprise est innovante et rompt avec le passé : les objets d’entreprise sont tous gérés suivant une modélisation commune sur trois vues (fonction, comportement et structure) et ceci indépendamment de leur nature (matérielle, temporelle, organisationnelle, etc.).

De ce fait, cette approche homogène des objets permet une conception et une maintenance du système d’information beaucoup plus aisées.

Ces objets pourront ensuite jouer différents rôles : processus, produit et ressource, ces trois rôles pouvant ensuite être raffinés en sous rôles si besoin est. Mais la structure, le comportement et les fonctions seront gérés de manière commune, indépendamment des rôles. Cette solution s’avère avantageuse et logique : par exemple la gestion d’un emploi du temps d’un salarié pourra intégrer des périodes de formation (le salarié, au travers du changement d’état de ses compétences, se trouve faire partie du produit du processus de formation), des périodes normales d’activité (il est alors considéré comme une ressource de l’activité), etc.

#### 6.7.3. Gestion originale de la notion de comportement

L’aspect comportemental n’est pas considéré en FBS-PPR comme étant une composante intrinsèque du modèle de produit (c’est par exemple le cas dans le modèle de produit de MOKA [DAI 00]). Le comportement de l’objet résulte de l’interaction entre un processus et cet objet.

Par voie de conséquence, toute création ou modification d’un objet (aspect dynamique, donc comportemental) nécessite la définition d’un processus associé. L’intérêt est une gestion plus cohérence des comportements (il est possible de définir autant de comportements que de processus associés) lorsque de nombreuses applications sont possibles pour un objet donné.

### 6.8. Conclusion

L’analyse des processus du cycle de vie produit est au cœur des problématiques industrielles. Elle met en évidence le rôle de la capitalisation et de la valorisation des connaissances.

L’attention portée aux processus métier, à leur finalité et à leur modélisation permet d’identifier les besoins et les productions de connaissance à commencer par les contenus du référentiel numérique.

La valorisation de ces contenus passe par une standardisation aussi poussée que possible, la difficulté principale résidant dans la représentation des connaissances permettant d'en garantir l'interprétation.

L'adoption et le déploiement d'un modèle de type FBS-PPR contribuent à l'analyse, la spécification et le suivi des éléments des processus d'entreprise. Ces éléments conceptuels constituent un support essentiel à la représentation de connaissances concernant l'ensemble des métiers du cycle de vie du produit en s'appuyant sur des vues, des sémantiques et des savoirs.

La vue processus a un rôle central structurant, et relie les vues produit et ressource. Du fait de la grande généricité de la brique FBS-PPR, les objets d'entreprises peuvent être décrits suivant le même formalisme indépendamment de leur rôle de produit (objet issus du processus) de ressource (moyen mis en œuvre dans un processus) ou encore de processus (organisation temporelle, spatiale et hiérarchique d'activités).

Pour tous ces objets sont différenciés les comportements réels et attendus. Ils permettent d'intégrer la prise en compte des aléas liés au cycle réel de ces objets dans l'entreprise et rendent plus aisée la définition d'indicateurs de performance adimensionnels.

Le modèle FBS-PPR a fait l'objet du développement d'un démonstrateur de système d'informations (cf. Figure 6.7.).

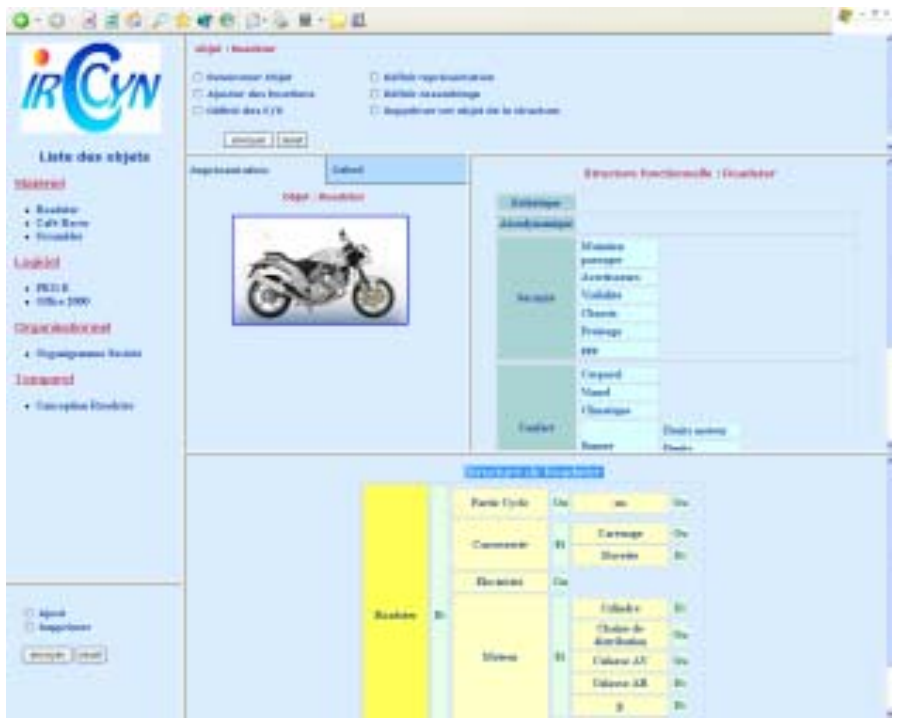

**Figure 6.7.** *Interface du démonstrateur FBS-PPR*

Il est structuré autour d'une architecture de type base de données, serveur et client via un navigateur (cf. Figure 6.8.). Il a permis de vérifier le bien fondé du concept FBS-PPR.

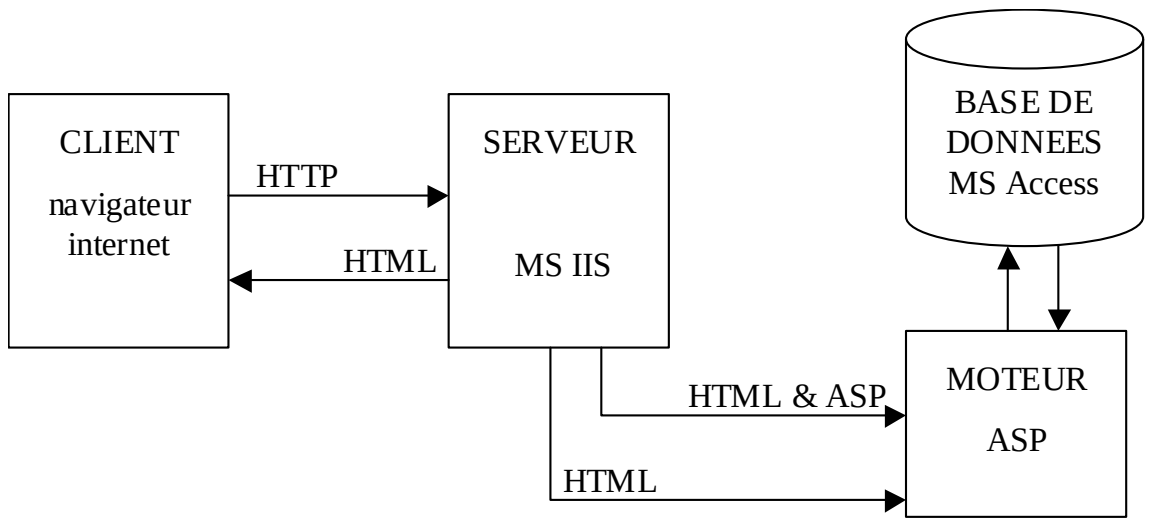

**Figure 6.8.** *Architecture du démonstrateur de système d'informations*

Il reste désormais à valider le modèle sur des cas d'application à plus grande échelle et en particulier d'évaluer les bénéfices réels pour les entreprises utilisatrices. Dans ce type de projets, les facteurs humains, économiques et de délais (d'autant que les bénéfices ne se font ressentir qu'a posteriori) sont généralement les principaux facteurs limitants.

## 6.9. Bibliographie